\documentclass[journal, 10pt]{IEEEtran}

\usepackage{cite}

\ifCLASSINFOpdf
  \usepackage[pdftex]{graphicx}
\else
  \usepackage[dvips]{graphicx}
  \graphicspath{{../eps/}}
  \DeclareGraphicsExtensions{.eps}
\fi

\usepackage{tikz}
\usepackage{tikzscale}
\usepackage{pgf}
\usepackage[T1]{fontenc} 

\usepackage{amsmath,amssymb,bm}
\usepackage[numbers,sort&compress]{natbib}
\usepackage{algorithmic}

\usepackage{array}
\ifCLASSOPTIONcompsoc
  \usepackage[caption=false,font=normalsize,labelfont=sf,textfont=sf]{subfig}
\else
  \usepackage[caption=false,font=footnotesize]{subfig}
\fi

\usepackage{stfloats}
\usepackage{url}

\usepackage{pgfplots}
\newcommand{\subparagraph}{}
\usepackage{titlesec}
\titlespacing*{\section}{1pt}{0.2\baselineskip}{0.2\baselineskip}
\titlespacing*{\subsection}{1pt}{0.2\baselineskip}{0.2\baselineskip}

\DeclareMathSizes{10}{9}{7}{5}

\begin{document}
\title{On The Uplink Throughput of Zero-Forcing in Cell-Free Massive MIMO with Coarse Quantization}
\IEEEaftertitletext{\vspace{-1.5\baselineskip}}

% use for special paper notices
%\IEEEspecialpapernotice{(Invited Paper)}

% author names and affiliations
% use a multiple column layout for up to three different
% affiliations
%\author{Author 1,\IEEEmembership{} Author 2\IEEEmembership{} and Author 3\IEEEmembership{}}

\author{Dick Maryopi,\IEEEmembership{} Manijeh Bashar\IEEEmembership{} and Alister Burr\IEEEmembership{}
\thanks{%Manuscript received May 20, 2018; revised May 30, 20018
%Copyright (c) 2015 IEEE. Personal use of this material is permitted. However, permission to use this material for any other purposes must be obtained from the IEEE by sending a request to pubs-permissions@ieee.org. 

The authors are with the Department of Electronic Engineering, University of York, Heslington, York, UK. email: dm1110@york.ac.uk. The paper was supported by Indonesia Endowment Fund for Education (LPDP) and in part by the European Horizon 2020 Programme under GA H2020-MSCA-ITN-2016-722788}}

\maketitle
\begin{abstract}
The recently proposed Cell-Free massive MIMO architecture is studied for the uplink. In contrast to most previous works, joint detection is performed using global CSI. Therefore, we study strategies for transferring CSI to the CPU taking into account the fronthaul capacity which limits CSI quantization. Two strategies for pilot-based CSI acquisition are considered: \emph{estimate-and-quantize} and \emph{quantize-and-estimate}. These are analysed using the Bussgang decomposition. For a given quantization constraint for the data and CSI the achievable rate per user with Zero-Forcing is determined. Numerical results show that quantize-and-estimate (the simpler strategy) is similar to or better than estimate-and-quantize at low resolution, especially for 1-bit.
\end{abstract}
% Note that keywords are not normally used for peerreview papers.
\begin{IEEEkeywords}
Cell-Free Massive MIMO, Fronthaul, Quantization, Bussgang, Channel Estimation.
\end{IEEEkeywords}

\IEEEpeerreviewmaketitle

%%===================================================================================================================
%%
%%===================================================================================================================
\section{Introduction}

\IEEEPARstart{T}{he} next generation of wireless networks (including 5G) will be required to provide a high capacity per user and per unit area due to the increasing number of users and the variety of applications expected in the near future. Cell-free massive Multiple Input Multiple Output (MIMO) has been gaining more attention recently as it has the potential to meet this demand \cite{7827017}. It can be regarded as a form of network MIMO which makes use of a large number of distributed antennas, referred to Access Points (AP), spread over a large coverage area. The term "cell-free" was motivated by the notion of blurring the role of cells so that all users can be served by all APs over the same resources using network MIMO techniques to avoid mutual interference. Because a large number of APs serve a smaller number of users, it still benefits from \emph{channel hardening} as in co-located massive MIMO \cite{5595728}.

Nevertheless, the joint transmission/detection in the most current cell-free system is based only on local Channel State Information (CSI). We identify this a limitation, since relying on local CSI at the APs restricts the feasible choice of processing to conjugate beamforming or Maximum Ratio Combining (MRC). Other forms of processing such as Zero Forcing (ZF) could be performed at Central Processing Unit (CPU), but would require additional CSI transfer via the fronthaul. However, CF-massive MIMO already faces the problem of high fronthaul load requirement. To address these issues, we study in this paper joint detection with global CSI at the CPU and the strategy of acquiring the required CSI. To deal with the growth of fronthaul load we assume a coarse quantization constraint, which is also of interest for the low-cost implementation of APs. We show that using appropriate CSI acquisition strategies, much improved detection techniques can be applied at the CPU resulting in a significant rate improvement in uplink. In \cite{7917284}, a performance improvement of ZF over MRC has also been shown in downlink where a sort of global CSI is used at the CPU for precoding. Nevertheless, they didn't address specifically the CSI acquisition schemes and the limited fronthaul capacity.

After a brief description of our system model we investigate two strategies of CSI acquisition. The first is called \emph{estimate-and-quantize} (EQ) where channel estimation is carried out at the AP. The channel estimate is quantized and then the quantized form is sent to the CPU. This is similar to the sharing of quantized CSI between the base stations in the coordinated multipoint (CoMP) scheme \cite{5706317}. As alternative we consider \emph{quantize-and-estimate} (QE), where the APs quantize the received pilot and send it to the CPU. From these quantized received pilots the CPU performs the channel estimation. Further, we compare their performance and their corresponding throughput for ZF detection. Surprisingly enough, the QE strategy, which is simpler for the implementation at the AP, has good performance and a significant performance improvement over EQ for 1-bit fronthaul resolution. Overall, the superiority of utilizing global CSI, even with coarse quantization, is shown to be significant compared to utilizing only local CSI with infinite resolution.

\emph{Notation}: Roman letters, lower-case boldface letters and upper-case boldface letters are used respectively to denote scalars, column vectors and matrices. The set of all complex and real $M\times N$ matrices are represented by $\mathbb{R}^{M \times N}$ and $\mathbb{C}^{M \times N}$ respectively. The real part and imaginary part of complex numbers are expressed respectively by $\operatorname{\mathbb{R}e}\{\cdot\}$ and $\operatorname{\mathbb{I}m}\{x\}$. By $\langle \cdot, \cdot\rangle$ we denote the inner product with $\Vert \cdot \Vert$ as its corresponding vector norm or Frobenius norm. The expectation of random variables is represented by $\mathbb{E}\{\cdot\}$. We denote the circularly complex Gaussian distribution with zero mean and unit variance by $\mathcal{CN}(0, 1)$. We use $\mathbf{I}_K$ for the $K\times K$ identity matrix and $\mathbf{1}_K$ for all-one vector of dimension $K$. For a vector $\mathbf{a}$, $\text{diag}(\mathbf{a})$ denotes a diagonal matrix with the diagonal elements taken from vector $\mathbf{a}$.

\section{System Model}

We consider the uplink transmission of a cell-free system \cite{7827017}, where we have $K$ single-antenna users (UEs) and $M$ single-antenna APs connected to a CPU by $M$ error-free fronthaul links. The main processing for these $M$ APs are virtualized at the CPU, where the communication between them occurs in baseband form. We assume that the fronthaul link connecting the $m$-th AP with the CPU can in practice transmit reliably at a maximum rate of $R_m$. 

\subsection{Channel Model}

The channel between the $k$-th user and the $m$-th AP is specified (as in \cite{7827017}) by
\begin{align}
g_{mk}=h_{mk} \beta_{mk}^{1/2}, \label{gmk}
\end{align}
where the coefficient $h_{mk}$ models the small-scale fading between the $k$-th user and the $m$-th AP with the assumption that it is i.i.d. $\sim \mathcal{C N}(0, 1)$. The large-scale fading is denoted by $\beta_{mk}$ which is likely to be different for each user $k$ and each AP $m$ due to the distributed configuration. The channel from all $K$ users to all $M$ APs can then be expressed as the element-wise product of small-scale fading matrix $\mathbf{H}\in\mathbb{C}^{M \times K}$ and large-scale fading matrix $\mathbf{D}\in\mathbb{R}^{M \times K}$ given by 
\begin{align}
\mathbf{G}=\mathbf{H} \odot \mathbf{D}^{1/2}, \text{ where } [\mathbf{H}]_{mk}=h_{mk}, [\mathbf{D}]_{mk}=\beta_{mk}.\label{GDS}
\end{align}

\subsection{Quantization Scheme}

To simplify our analysis, we consider fronthaul links with $R_m=R$ bits, $\forall m \in \{1,\dots,M\}$, corresponding to the quantization level $L=2^{R}$. Therefore, we apply an $L$-level scalar quantizer $Q$ at each AP as an interface to the fronthaul with 
\begin{align}
Q(x)= \sum_{l=0}^{L-1} q_l T_l(x), \label{QFormula}
\end{align}
where $T_l(\mathbf{x})$ is equal 1 for $x_l<x\leq x_{l+1}$ and 0 otherwise. We consider $Q$ as a uniform quantizer with a fixed step size $\Delta=x_{l+1}-x_l$ and a reconstruction value $q_l=(l-\frac{L-1}{2})\Delta$. For a complex-valued signal $x\in \mathbb{C}$ we quantize the real and imaginary part separately. In this case, whenever we have $x_l<\operatorname{\mathbb{R}e}\{x\}\leq x_{l+1}$ and $x_{l'}<\operatorname{\mathbb{I}m}\{x\}\leq x_{l'+1}$ for $(l, l') \in\{0, \dots, L-1\}$, we obtain
\begin{align}
x_q=Q(x)&=Q(\operatorname{\mathbb{R}e}\{x\})+ iQ(\operatorname{\mathbb{I}m}\{x\})\\ 
&=q_l^{R}+iq_{l'}^{I},
\end{align}
where $q_l^{R}$ and $q_{l'}^{I}$ are respectively the reconstruction values of the real and imaginary part with the pair $(q_l^{R}, q_{l'}^{I}) \in \{q_0^{R}, \dots, q_{L-1}^{R}\} \times \{q_0^{I}, \dots, q_{L-1}^{I}\}$. Moreover, the quantization operation should apply elementwise for a vector valued input. We assume that the large scale fading $\beta_{mk}$ is relatively constant over a long period and known at the APs. Thus, we can scale the input-output signal of the quantizer according to $\beta_{mk}$ and approximate the normalised input as normally distributed. 

The function $Q$ is the scalar quantization process, which is particularly nonlinear for small $L$. To analyse it, we use the Bussgang decomposition \cite{Bussgang52}. Accordingly, for a nonlinear function $Q(x)$ we can write it as 
\begin{align}
x_q=Q(x)=\alpha_q x + d. \label{BusgangAlpha}
\end{align}
The distortion term $d$ is uncorrelated to the input signal $x$. The linear factor $\alpha_q$ depends on the characteristic of the quantizer $Q$ and the distribution $f(x)$ of the input signal $x$ given by \cite{Bussgang52, 5560739}
\begin{align}
\alpha_{q}&=\frac{1}{P_x}\int_{x} x Q^*(x)f(x) dx \nonumber\\
&=\frac{1}{P_x}\sum_{l=0}^{L-1} q_l \int_{x_l}^{x_{l+1}} xf(x) dx, 
\end{align}
where $P_x=\mathbb{E}\{\vert x\vert^2\}$ is the power of $x$. As shown in \cite{VTC18BBM}, for normally distributed input and uniform quantizer it can be expressed in closed form as a function of $\Delta$ and $L$  
\begin{align}
\alpha_q=\frac{\Delta}{\sqrt{2\pi}}\left(1+2\sum_{l=1}^{L/2-1}\text{exp}(-l^2\Delta^2)\right). \label{Alpha}
\end{align}
Further, we define the power ratio of the input $x$ and the output $x_q$ in terms of $\Delta$ and $L$ as given in \cite{VTC18BBM} by
\begin{align}
\lambda_q=\frac{\mathbb{E}\{\vert x_q\vert^2\}}{\mathbb{E}\{\vert x\vert^2\}}
&=\frac{1}{P_x}\int_{x} \vert Q(x)\vert^2 f(x) dx\\
&=\frac{1}{P_x}\sum_{l=0}^{L-1} q_l^2 \int_{x_i}^{x_{l+1}} f(x) dx\\
&=\Delta^2\left(\frac{1}{4}+4\sum_{l=1}^{L/2-1} l(1-\Phi(l\Delta))\right), \label{AlphaRatio}
\end{align}
where $\Phi$ is the Gaussian cumulative distribution function. We choose here the step size $\Delta$ that maximizes the Signal to Distortion Noise Ratio (SDNR) at the output of the quantizer defined as
\begin{align}
\text{SDNR}=\mathbb{E}\{\vert \alpha_q x\vert^2\}/\mathbb{E}\{\vert d\vert^2\}.
\end{align}
From (\ref{BusgangAlpha}) and (\ref{AlphaRatio}) the power of the distortion is given by
\begin{align}
\mathbb{E}\{\vert d\vert^2\}&=\mathbb{E}\{\vert x_q-\alpha x\vert^2\}
=(\lambda_q-\alpha_q^2)\mathbb{E}\{\vert x\vert^2\}. \label{outputPower}
\end{align}
Using equations (\ref{Alpha}) and (\ref{AlphaRatio}) we characterize the Bussgang decomposition such that it is directly related to the parameter $\Delta$ and $L$. This will be useful for the analysis and numerical evaluation of the quantization process.

\section{CSI Acquisition Strategies}
The CSI is acquired based on the estimation of known pilots transmitted by the users. In this case, the $k$-th user transmits $\sqrt{\tau_p}\bm{\varphi}_k$ as its pilot, where a specific random sequence $\bm{\varphi}_k \in\mathbb{C}^{\tau_p\times 1}$ is taken from an orthonormal basis with $\vert\langle\bm{\varphi}_k, \bm{\varphi}_k'\rangle\vert=\delta_{kk'}$ and $\Vert\bm{\varphi}_k\Vert ^2=1$. The sequence length $\tau_p$ is assumed to be less than or equal to the coherence interval $\tau_c$. The $m$-th AP observes the received pilot $\mathbf{y}_{m}$ from all $K$ users as
\begin{align} \label{receivePilot}
\mathbf{y}_{p,m}= \sqrt{\tau_p\rho_p}\sum_{k=1}^{K} g_{mk}\bm{\varphi}_k + \mathbf{w}_p,
\end{align}
where $\rho_p$ is the transmit SNR of the pilot and $\mathbf{w}_p\sim \mathcal{C N}(0, \mathbf{I}_K)$ is an additive noise vector with zero mean and identity covariance. To ensure that all pilots are orthogonal for all $K$ users, one should only allow $K\leq \tau_p$ users who transmit their pilots simultaneously. In this case, the transmitted pilots satisfy
\begin{align}
\Theta^H\Theta=\tau_p\rho_p \mathbf{I}_K,\text{ where } \Theta=\sqrt{\tau_p\rho_p}[\bm{\varphi}_1, \dots,\bm{\varphi}_K].
\end{align}

In the ideal case of perfect fronthaul \cite{7827017} the channel $g_{mk}$ can be estimated at the AP and sent to the CPU which then has the global CSI. In this case, the received pilot $\mathbf{y}_{p,m}$ at the $m$-th AP is projected onto $\bm{\varphi}_k^H$ giving:
\begin{align}
r_{p,mk}&=\bm{\varphi}_k^H\mathbf{y}_{p,m}\nonumber\\
&=\sqrt{\tau_p\rho_p}g_{mk}+\sqrt{\tau_p\rho_p}\sum_{k'\neq k}^{K} g_{mk'}\bm{\varphi}_k^H\varphi_k' + \bm{\varphi}_k^H \mathbf{w}_p. \label{projectPilot}
\end{align}
To obtain the estimate of $g_{mk}$ we use the Linear Minimum Mean Squared Error (LMMSE) estimator given by
\begin{align}
\hat{g}_{mk}= c_{mk}r_{p,mk}. \label{EstimatorIdeal}
\end{align}
We choose $c_{mk}$ that minimizes the Mean Squared Error (MSE) 
\begin{align}
\epsilon_{mk}&=\mathbb{E}\{\vert g_{mk}-\hat{g}_{mk}\vert^2\}. \label{MSEgmk}
\end{align}

The unique minimum is obtained by taking the derivative of $\epsilon_{mk}$ and setting it equal to zero giving
\begin{align}
c_{mk}\!&=\!\frac{\operatorname{\mathbb{R}e}\{\mathbb{E}\{r_{p,mk}^*g_{mk}\}\}}{\mathbb{E}\{\vert r_{p,mk}\vert^2\}}\nonumber\\
\!&=\!\frac{\sqrt{\tau_p \rho_p} \beta_{mk}}{\tau_p\rho_p\sum_{k'=1}^{K}\beta_{mk'}\vert\bm{\varphi}_k^H\bm{\varphi}_{k'}\vert^2+1},
\end{align}
where the last equation follows from (\ref{projectPilot}). With the optimal coefficient $c_{mk}$ the minimum mean squared error is then given by
\begin{flalign}
\epsilon_{mk}=\mathbb{E}\{\vert g_{mk}\vert^2\}-\frac{(\operatorname{\mathbb{R}e}\{\mathbb{E}\{r_{p,mk}^*g_{mk}\}\})^2}{\mathbb{E}\{\vert r_{p,mk}\vert^2\}}=\beta_{mk}-\gamma_{mk}, \label{MSEemk}
\end{flalign}
where we use $\gamma_{mk}$ to denote the mean squared of the channel estimate given by 
\begin{align}
\gamma_{mk}\triangleq\mathbb{E}\{\vert\hat{g}_{mk}\vert^2\}
&=c_{mk}^2\mathbb{E}\{\vert r_{p,mk}\vert^2\}\nonumber\\
&=c_{mk}\operatorname{\mathbb{R}e}\{\mathbb{E}\{r_{p,mk}^*g_{mk}\}\}\nonumber\\
%&=c_{mk}\sqrt{\tau\rho_p}\beta_{mk}\nonumber\\
&=\frac{\tau_p \rho_p \beta_{mk}^2}{\tau_p\rho_p\sum_{k'=1}^{K}\beta_{mk'}\vert\bm{\varphi}_k^H\bm{\varphi}_{k'}\vert^2+1}.
\end{align}

We suppose that the CSI is transferred to the CPU in the same time frame as the uplink data. To possibly maximize the rate, the same proportion of power is allocated to pilot and to data as in the training-based scheme of general MIMO system \cite{1193803}. Let $\rho$ and $\rho_u$ denote the total transmit SNR and the transmit Signal to Noise Ratio (SNR) for the uplink data respectively, the power allocation for pilot of length $\tau_p$ and for data of length $\tau_u$ follows
\begin{align}
\rho_u \tau_u= \frac{\rho \tau_c}{2} \text{ and } \rho_p \tau_p=\frac{\rho \tau_c}{2}, \text{ where } \tau_c=\tau_p+\tau_u.
\end{align}

\subsection{Estimate-and-Quantize}

In this scheme we estimate the channel coefficient $g_{mk}$ first as given in (\ref{EstimatorIdeal}). So that it may be sent via limited fronthaul to the CPU, the estimated channel $\hat{g}_{mk}$ is quantized at each AP. Because we send the quantized version $\hat{g}_{mk}^{eq}$ to the CPU, the amount of CSI overhead resulted by this scheme is proportional to the number of users $K$. For symbol frame of length $\tau_c$ the portion of CSI overhead is then $K/\tau_c$. After transferring via the fronthaul the CPU receives $\hat{g}_{mk}^{eq}$, which can be decomposed by Bussgang as
\begin{align}
\hat{g}_{mk}^{eq}=Q(\hat{g}_{mk})=\alpha_{eq}\hat{g}_{mk}+d_{eq}. \label{BussgangQC}
\end{align}  
The mean squared error after quantization is given by
\begin{align}
\epsilon_{mk}^{eq}&=\mathbb{E}\{\vert g_{mk}-\hat{g}_{mk}^{eq}\vert^2\}\\
&=\mathbb{E}\{\vert g_{mk}\vert^2\}+\mathbb{E}\{\vert \hat{g}_{mk}^{eq}\vert^2\}-2\operatorname{\mathbb{R}e}\{\mathbb{E}\{g_{mk}^*\hat{g}_{mk}^{eq}\}\}.
\end{align}
We can apply (\ref{BussgangQC}) to express
\begin{align}
\mathbb{E}\{g_{mk}^*\hat{g}_{mk}^{eq}\}&=\alpha_{eq}\mathbb{E}\{ g_{mk}^*\hat{g}_{mk}\}+\mathbb{E}\{g_{mk}^*d_{eq}\}\nonumber\\
&=\alpha_{eq}\mathbb{E}\{ g_{mk}^*\hat{g}_{mk}\},
\end{align}
where the second term vanishes because $g*_{mk}$ is uncorrelated with $d_eq$. This follows because $\mathbb{E}\{\hat{g}_{mk}d_{eq}\}=0$ and our use of a linear MMSE estimator means that the estimation error prior to quantization is also uncorrelated with $\hat{g}_mk$ and hence also with $d_eq$. We then obtain
\begin{align}
\epsilon_{mk}^{eq}&=\mathbb{E}\{\vert g_{mk}\vert^2\}+\lambda_{eq}\mathbb{E}\{\vert\hat{g}_{mk}\vert^2\}- 2\alpha_{eq}\operatorname{\mathbb{R}e}\{\mathbb{E}\{g_{mk}^* \hat{g}_{mk}\}\}.\nonumber\\
&=\mathbb{E}\{\vert g_{mk}\vert^2\}+\lambda_{eq}\gamma_{mk}- 2\alpha_{eq}\gamma_{mk}\nonumber\\
&=\beta_{mk}-(2\alpha_{eq}-\lambda_{eq})\gamma_{mk}. \label{MSEQCAnal}
\end{align}
Note that in the practical implementation of this scheme the channel estimation does not have to be performed at low resolution: the channel can be estimated at the AP at high precision, in the same way as CSI quantization in the CoMP scenario, and the estimate subsequently quantized at a lower resolution, in order to reduce the fronthaul load.

\subsection{Quantize-and-Estimate}

Unlike the previous scheme, here we quantize the pilot first and send it to the CPU to estimate $g_{mk}$. In this case, at the CPU we have the quantized received pilots which once again may be decomposed using the Bussgang decomposition as
\begin{align}
\mathbf{y}_{p,m}^{q}&=Q(\mathbf{y}_{p,m})=\alpha_{qe} \mathbf{y}_{p,m}+\mathbf{d}_{qe}.
\end{align}
The noisy quantized observation at the CPU is given as
\begin{align}
r_{p,mk}^q\!=\!\bm{\varphi}_k^H \mathbf{y}_{p,m}^q
\!=\!\alpha_{qe} \bm{\varphi}_k^H \mathbf{y}_{p,m}+ \bm{\varphi}_k^H \mathbf{d}_{qe}\nonumber\\
\!=\!\alpha_{qe} r_{p,mk}+  \bm{\varphi}_k^H \mathbf{d}_{qe}. \label{receivedQP}
\end{align}
We then apply the LMMSE estimator to obtain the quantize-and-estimate channel coefficient $\hat{g}_{mk}^{qe}$ given by
\begin{align}
\hat{g}_{mk}^{qe}\!&=\!c_{mk}^{qe} r_{p,mk}^q,
\end{align}
where we choose $c_{mk}^{qe}$ that minimizes the MSE $\mathbb{E}\{\vert g_{mk}-\hat{g}_{mk}^{qe}\vert^2\}$. As derived in Appendix \ref{SecondAppendix} the coefficient $c_{mk}^{qe}$ is given by
\begin{align}\label{QuantEstimator}
c_{mk}^{qe}\!&=\!c_{mk}\frac{\alpha_{qe}a_{mk}}{\alpha_{qe}^2a_{mk}+(\lambda_{qe}-\alpha_{qe}^2)b_m}, \text{ where } \\ 
a_{mk}\!&\triangleq\!\tau_p\rho_p\sum_{k'=1}^{K}\beta_{mk'}\vert\bm{\varphi}_k^H\bm{\varphi}_{k'}\vert^2+1, \text{ and }  
b_m\!\triangleq\!\rho_p\sum_{k=1}^{K}\beta_{mk}+1.  \nonumber
\end{align}
We then also obtain the MSE for the QE scheme expressed as (see Appendix \ref{SecondAppendix})
\begin{alignat}{1}
\epsilon_{mk}^{qe}&=\mathbb{E}\{\vert g_{mk}-\hat{g}_{mk}^{qe}\vert^2\}\nonumber\\ 
&=\beta_{mk}-\left(\frac{\alpha_{qe}^2 a_{mk}}{\alpha_{qe}^2a _{mk}+(\lambda_{qe}-\alpha_{qe}^2)b_m}\right)\gamma_{mk}. \label{MSEQPAnal}
\end{alignat}

Because we quantize the received pilot in this scheme, the amount of the resulted CSI overhead is proportional to the length of pilot $\tau_p$ which doesn't scale directly with the number of users. However, in the case of orthogonal pilots the EQ and QE scheme have the same amount of CSI overhead. In terms of complexity at AP, this scheme apparently has lower complexity than EQ scheme bacause no estimator and only a single quantizer are needed.

\section{The Achievable Rates with Coarse Quantization}

The uplink data received at all $M$ APs may be described by
\begin{align}
\mathbf{y}_d= \sqrt{\rho_u} \mathbf{G}\mathbf{x}_d+\mathbf{w}_d,
\end{align}
where $\mathbf{x}_d \in \mathbb{C}^K$ is the transmitted data from all $K$ users, $\mathbf{G}$ is the channel matrix defined in (\ref{GDS}) and $\mathbf{w}_d\sim \mathcal{C N}(0, \mathbf{I}_M)$ is an additive noise vector. After quantization and transmission via the fronthaul the CPU obtains the data signal $\mathbf{r}_d$, which can also be decomposed as
\begin{align}
\mathbf{r}_d=Q(\mathbf{y}_d)
&=\alpha_{qd}\mathbf{y}_d+\mathbf{d}_{qd}\\
&=\sqrt{\rho_u}\alpha_{qd} \mathbf{G}\mathbf{x}_d+\alpha_{qd}\mathbf{w}_d+\mathbf{d}_{qd}\nonumber\\
&=\sqrt{\rho_u}\alpha_{qd}\hat{\mathbf{G}}\mathbf{x}_d+\sqrt{\rho_u}\alpha_{qd}\tilde{\mathbf{G}}\mathbf{x}_d+\alpha_{qd}\mathbf{w}_d+\mathbf{d}_{qd},\nonumber
\end{align}
where $\tilde{\mathbf{G}}$ is the channel estimation error including the quantization error. Further, we treat $\hat{\mathbf{G}}$ as the true channel and treat the second term and so forth as an effective noise $\mathbf{z}$ such that
\begin{flalign}
\mathbf{r}_d=\sqrt{\rho_u}\alpha_{qd} \hat{\mathbf{G}}\mathbf{x}_d+\mathbf{z}\label{receivedData}
\end{flalign}
%---------------------------------------
To detect the transmitted data we can use a ZF detection matrix $\bar{\mathbf{A}}^H=(\hat{\mathbf{G}}^H\hat{\mathbf{G}})^{-1}\hat{\mathbf{G}}^H$, with $\bar{\mathbf{A}}^H\hat{\mathbf{G}}=\mathbf{I}_K$. We then obtain the estimated data as
\begin{align}
\mathbf{\hat{x}}_d&=\bar{\mathbf{A}}^H\mathbf{r}_d=\sqrt{\rho_u}\alpha_{qd}\mathbf{x}_d+\bar{\mathbf{A}}^H\mathbf{z},
\end{align}
such that the SINR for the $k$-th user is given by
\begin{align}\label{directSINR}
\operatorname{SINR}_k^{ZF}=\frac{\rho_u\alpha_{qd}^2}{\left[\mathbb{E}\{ \bar{\mathbf{A}}^H\mathbf{z}\mathbf{z}^H\bar{\mathbf{A}} \}\right]_{k,k}}.
\end{align}
%---------------------------------------
However, due to the nature of the matrix $\mathbf{G}$ in the case of distributed massive MIMO, which tends to have independent large scale fading coefficients, the closed form expression of Signal to Interference Noise Ratio (SINR) for ZF is intractable. To obtain the SINR expression for our quantized CF massive MIMO we follow the approximation derived in \cite{7868985}.
We apply $\hat{\mathbf{G}}$ in our zero forcing detector to detect the data from (\ref{receivedData}) and apply a filter $\mathbf{\Lambda}^{-1/2}$ to $\mathbf{r}_d$ to whiten $\mathbf{z}$, such that we have a ZF detector matrix
\begin{align}
\mathbf{A}^H&=(\hat{\mathbf{G}}^H\mathbf{\Lambda}^{-1}\hat{\mathbf{G}})^{-1}\hat{\mathbf{G}}^H\mathbf{\Lambda}^{-1/2},\text{ where } \\
\mathbf{\Lambda}&=\mathbb{E}\{\mathbf{z}\mathbf{z}^H\} \text{ and } \mathbf{A}^H \mathbf{\Lambda}^{-1/2}\hat{\mathbf{G}}=\mathbf{I}_K.\label{Lambda}
\end{align}
After detection we obtain
\begin{align}
\hat{\mathbf{x}}_d\!&=\!\mathbf{A}^H \!\mathbf{\Lambda}^{-1/2}\mathbf{r}_d \nonumber\\
\!&=\!\sqrt{\rho_u}\alpha_{qd}\mathbf{x}_d\!+\!(\hat{\mathbf{G}}^H \!\mathbf{\Lambda}^{-1}\hat{\mathbf{G}})^{-1}\hat{\mathbf{G}}^H \!\mathbf{\Lambda}^{-1}\mathbf{z}.
\end{align}
The instantaneous SINR (i.e. the SINR for a specific realization of $\mathbf{z}$) for the $k$-th user can then be expressed as
\begin{flalign} \label{SINRlong}
\operatorname{SINR}_k^{ZF}\!=\!\frac{\rho_u \alpha_{qd}^2 }{\left[\!(\hat{\mathbf{G}}^H\!\mathbf{\Lambda}\!^{-1}\hat{\mathbf{G}})^{-1}\hat{\mathbf{G}}^H\!\mathbf{\Lambda}\!^{-1}\mathbf{z}\mathbf{z}^H\!\mathbf{\Lambda}\!^{-1}\hat{\mathbf{G}}(\hat{\mathbf{G}}^H\!\mathbf{\Lambda}\!^{-1}\hat{\mathbf{G}})^{-1}\!\right]_{k,k}}
\end{flalign}
Following \cite{7868985} we may approximate $\mathbf{z}\mathbf{z}^H$ in (\ref{SINRlong}) by its expectation $\mathbf{\Lambda}$ such that it remains
\begin{align} \label{SINRshort}
\operatorname{SINR}_k^{ZF} \overset{(\ref{Lambda})}{\approx} \frac{\rho_u \alpha_{qd}^2 }{\left[(\hat{\mathbf{G}}^H\mathbf{\Lambda}^{-1}\hat{\mathbf{G}})^{-1}\right]_{k,k}}
\end{align}
In this way, we can express the $\text{SINR}_k^{ZF}$ as \cite{7868985}
\begin{align}\label{approxSINR}
\operatorname{SINR}_k^{ZF} & \approx \rho_u\alpha_{qd}^2 \left(\frac{M-K+1}{M}\right) \hat{g}_k^H \mathbf{\Lambda}^{-1}\hat{g}_k.
\end{align}
Due to the independent realization of the additive noise and estimation error at each AP we may assume that the effective noise $\mathbf{z}$ is uncorrelated over $M$ APs. Thus, the matrix $\mathbf{\Lambda}$ is a diagonal matrix given by
\begin{align}
\mathbf{\Lambda}\!=\!\operatorname{diag}\{\Lambda_1,..., \Lambda_M\}\text{ and } 
\Lambda_m &\!=\!\sigma_{d_{qd}}^2\!+\!\alpha_{qd}^2\sigma_n^2\!+\!\rho_u\alpha_{qd}^2 \!\sum_{k=1}^K \! \epsilon_{mk}^q, \nonumber
\end{align}
where $\sigma_{d_{qd}}^2$ is the distortion variance resulted from quantizing data, $\sigma_n^2$ is the noise variance and $\epsilon_{mk}^q \in\{\epsilon_{mk}^{eq}, \epsilon_{mk}^{qe} \}$ is the estimation error from (\ref{MSEQCAnal}) or (\ref{MSEQPAnal}) depending on the scheme. The achievable rate per user in the uplink is then given by
\begin{align}
R_{u,k}^{ZF}=\operatorname{log}_2\left(1+\operatorname{SINR}_k^{ZF}\right). \label{RKzf}
\vspace{-10pt}
\end{align}

%==================================================================
\section{Numerical Results}

In the following, we provide some numerical results for the considered schemes above. We do simulations with system parameters similar to \cite{7827017} where there are $M=200$ APs and $K=20$ users distributed uniformly in an area of $1\times 1 \text{ km}^2$. We assume that this simulation area is wrapped around to avoid the boundary effects. For the channel $g_{mk}$ given in (\ref{gmk}) we model the large scale fading
$\beta_{mk}=\text{PL}_{mk}\cdot 10^{(\sigma_{sh}z_{mk})/10}$,
where the factor $10^{(\sigma_{sh}z_{mk})/10}$ is the uncorrelated shadowing with the standard deviation $\sigma_{sh}= 8 \text{ dB}$ and $z_{mk}\sim \mathcal{N}(0, 1)$. The path loss coefficient follows the three-slope model according to 
\begin{flalign}
&\text{PL}_{mk}\!=\!
\begin{cases}
\!-\!\mathcal{L}\!-\!35\text{log}_{10}(d_{mk}), d_{mk} \! >\! d_1 \\
\!-\!\mathcal{L}\!-\!15\text{log}_{10}(d_1)\!-\!20\text{log}_{10}(d_{mk}), d_0\!<\!d_{mk}\!\leq \!d_1 \\
\!-\!\mathcal{L}\!-\!15\text{log}_{10}(d_1)\!-\!20\text{log}_{10}(d_0), d_{mk} \!\leq \!d_0, 
\end{cases} \nonumber
\end{flalign}
where $d_{mk}$ is the distance between the $m$-th AP and the $k$-th user, $d_0=0.01$km, $d_1=0.05$km, and
\begin{align}
\mathcal{L}&\triangleq 46.3+33.9 \operatorname{log}_{10}(f)-13.83 \operatorname{log}_{10}(h_{AP})\nonumber\\
&-(1.1 \operatorname{log}_{10}(f)-0.7)h_u+(1.56 \operatorname{log}_{10}(f)-0.8).
\end{align}
We choose the carrier frequency $f=1.9 \text{ GHz}$, the AP antenna height $h_{AP}=15\text{m}$ and the user antenna height $h_u=1.65\text{m}$. In our simulation the normalized transmit SNRs $\rho_u$ and $\rho_p$ are defined as the transmit power divided by the noise power which is $B\times k_b\times T_0 \times \text{noise figure}$. We suppose that the bandwidth $B=20\text{ MHz}$, the Boltzmann constant $k_b=1.381 \times 10^{-23}$, the noise temperature $T_0=290$ Kelvin and the noise figure $=9$ dB. To make a fair comparison, our simulation considers the orthogonal case with $\tau_p=K$ where EQ and QE scheme spend the same length of CSI overhead. We allocate 10\% of symbols for acquiring CSI where $\tau_p=20$ symbols are spent for the pilot from overall $\tau_c=200$. 

\begin{figure}[ht!]
\centering
\includegraphics[width=0.9\columnwidth]{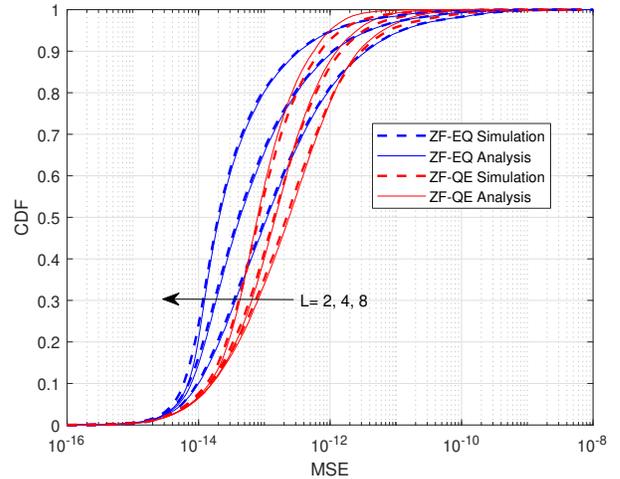}
\vspace{-0.1in}
\caption{The cumulative distribution of the channel estimation MSE $\epsilon_{mk}^q$ for the schemes estimate-and-quantize (EQ) in (\ref{MSEQCAnal}) and quantize-and-estimate (QE) in (\ref{MSEQPAnal}) with $K=20$, $M=200$ and Transmit Power $=0$ dBW}
\label{fig1}
\end{figure}

\begin{figure}[ht!]
\centering
\includegraphics[width=0.9\columnwidth]{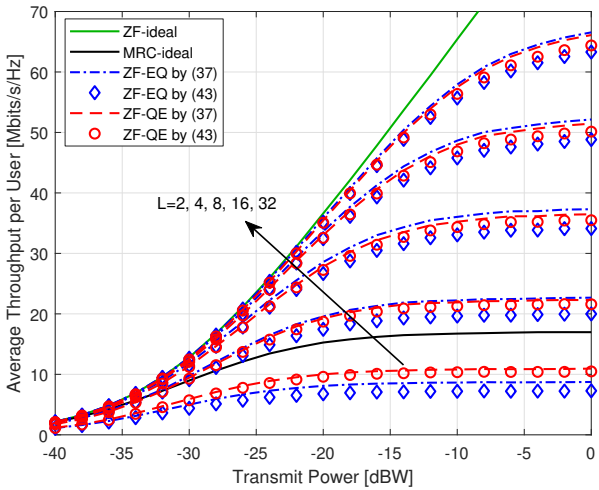}
\vspace{-0.1in}
\caption{The average per user throughput for different number of quantization level $L$, transmit power and CSI acquisition schemes for $K=20$ and $M=200$.}
\label{fig2}
\end{figure}

In Fig \ref{fig1}, we first validate with simulations our analytical MSE approximations which are obtained in (\ref{MSEQCAnal}) and (\ref{MSEQPAnal}) using Bussgang decomposition. Note that in our simulation setup the large scale fading has very small value up to -17 order of magnitude. This boils down to very small channel gain and to very small typical value of MSE. It is shown in Fig. \ref{fig1} that our analyses for both strategies are quite close to simulations especially for small $L$ and high transmit power. In at least $80\%$ of cases (those with the lower MSE) the QE scheme gives a poorer MSE than EQ and for $L>2$ this proportion increases. However, it is the larger channel estimate errors that have stronger influence on the rate. Using the corresponding channel estimation errors we then evaluate the average achievable rates per user given in (\ref{RKzf}). In this case, we compare their performance in terms of their per-user net throughput defined as
\begin{align}
S_{u,k}^{ZF}&\triangleq B\frac{1-\tau_p/\tau_c}{2}R_{u,k}^{ZF},
\end{align}
where the CSI overhead is taken into account by the term $1-\tau_p/\tau_c$. As shown in Fig. \ref{fig2} the QE scheme achieves higher throughput than the EQ scheme for small $L$ over the whole range of transmit power. The performance gap is decreasing as we increase the quantization level. For small $L$, the achievable rates computed by our approximation (\ref{approxSINR}) has only relatively small deviation from the rate computed by (\ref{directSINR}). It can also clearly be observed that ZF with low quantization level $L=4$ can already outperform MRC even with infinite quantization precision. This demonstrates the great improvement resulting from having global CSI available at the CPU. With $L=32=5$-bits we are about $5$ dB away from ZF with ideal fronthaul to reach $60$ Mbits/s/Hz average throughput per user. Meanwhile, the trade off between the increasing throughput and the resulting latency due to CSI overhead is left for future works.

\section{Conclusion}

This paper shows the benefit of having global CSI at the CPU for the uplink of cell-free massive MIMO.
We have established the MSE expression of CSI-acquisition strategies and compared their performance. We have presented their corresponding average throughput for ZF detection. In this case, the low-complexity scheme ZF-QE outperforms ZF-EQ at low resolution especially for 1-bit. 
%==================================================================
\appendices

\section{Derivation of Equation (\ref{QuantEstimator}) and (\ref{MSEQPAnal})}

\label{SecondAppendix}
For the estimator in (\ref{QuantEstimator}) we have $c_{mk}^{qe}$ that minimizes the MSE given by
\begin{align}
c_{mk}^{qe}=\frac{\operatorname{\mathbb{R}e}\{\mathbb{E}\{r_{p,mk}^{q^*}g_{mk}\}\}}{\mathbb{E}\{\vert r_{p,mk}^q\vert^2\}}.
\end{align}
From (\ref{receivedQP}) we can express the numerator of $c_{mk}^{qe}$ as
\begin{align}
\operatorname{\mathbb{R}e}\{\mathbb{E}\{r_{p,mk}^{q^*}g_{mk}\}\}&=\alpha_{qe} \operatorname{\mathbb{R}e}\{\mathbb{E}\{r_{p,mk}^* g_{mk}\}\}+\operatorname{\mathbb{R}e}\{\mathbb{E}\{\bm{\varphi}_k^H \mathbf{d}_{qe} g_{mk}\}\}\nonumber\\
&=\alpha_{qe} \sqrt{\tau_p\rho_p} \beta_{mk}, \label{numerMSEQP}
\end{align}
where the second term vanishes due to uncorrelation. Likewise we can express the denominator as
\begin{align}
\mathbb{E}\{\vert r_{p,mk}^q\vert^2\}&=\alpha_{qe}^2\mathbb{E}\{\vert r_{p,mk}\vert^2\}+\mathbb{E}\{\vert \bm{\varphi}_k^H \mathbf{d}_{qe}\vert^2\},\label{denomMSEQP}
\end{align}
where the first term is given by
\begin{align}
\alpha_{qe}^2\mathbb{E}\{\vert r_{p,mk}\vert^2\}=\alpha_{qe}^2\left(\tau_p\rho_p\sum_{k'=1}^{K}\beta_{mk'}\vert\bm{\varphi}_k^H\bm{\varphi}_{k'}\vert^2+1\right)
\end{align}
and the second term is given by
\begin{align}
\mathbb{E}\{\vert \bm{\varphi}_k^H \mathbf{d}_{qe}\vert^2\}&= \Vert\bm{\varphi}_k^H\Vert^2 \mathbb{E}\{\vert \mathbf{d}_{qe}\vert^2\}
\overset{(\ref{outputPower})}{=}(\lambda_{qe}-\alpha_{qe}^2)\mathbb{E}\{\vert \mathbf{y}_{p,m}\vert^2\}\nonumber\\
&=(\lambda_{qe}-\alpha_{qe}^2)\left(\rho_p\sum_{k=1}^{K}\beta_{mk}+1\right).
\end{align}
Let $a_{mk}$ and $b_m$ denote the following expressions
\begin{align}
a_{mk}\!&\triangleq\!\tau_p\rho_p\sum_{k'=1}^{K}\beta_{mk'}\vert\bm{\varphi}_k^H\bm{\varphi}_{k'}\vert^2+1, \text{ and }  
b_m\!\triangleq\!\rho_p\sum_{k=1}^{K}\beta_{mk}+1,  \nonumber
\end{align}
then we obtain
\begin{align}
c_{mk}^{qe}&=\frac{\alpha_{qe} \sqrt{\tau_p\rho_p} \beta_{mk}}{\alpha_{qe}a_{mk}}\frac{\alpha_{qe}a_{mk}}{\alpha_{qe}^2a_{mk}+(\lambda_{qe}-\alpha_{qe}^2)b_m}\nonumber\\
&=c_{mk}\frac{\alpha_{qe}a_{mk}}{\alpha_{qe}^2a_{mk}+(\lambda_{qe}-\alpha_{qe}^2)b_m}.
\end{align}
Further, we have the MSE given by
\begin{align}
\epsilon_{mk}^{qe}=\mathbb{E}\{\vert g_{mk}\vert^2\}-\frac{(\mathbb{E}\{r_{p,mk}^{q^*}g_{mk}\})^2}{\mathbb{E}\{\vert r_{p,mk}^q\vert^2\}},
\end{align}
where the second term can also be expressed as
\begin{align}
\gamma_{mk}^{qe}&=\frac{\alpha_{qe}^2\tau \rho_p \beta_{mk}^2}{\alpha_{qe}^2a_{mk}}\frac{\alpha_{qe}^2a_{mk}}{\alpha_{qe}^2a_{mk}+(\lambda_{qe}-\alpha_{qe}^2)b_m}\nonumber\\
&=\gamma_{mk}\frac{\alpha_{qe}^2 a_{mk}}{\alpha_{qe}^2a _{mk}+(\lambda_{qe}-\alpha_{qe}^2)b_m}.
\end{align}
%===================================================================================================================
\bibliographystyle{IEEEtran}

{\footnotesize
\bibliography{References}}

\nocite{*}

\IEEEpeerreviewmaketitle

\end{document}